\documentstyle[preprint,aps]{revtex}
\begin{document}
\preprint{DAMTP/R-94/3}
\draft
\tighten
\title{\large\bf Pair production of black holes in a $U(1) \otimes U(1)$
theory}
\author{S.F. Ross\footnote{E-mail: S.F.Ross@amtp.cam.ac.uk}}
\address{Department of Applied Mathematics and Theoretical Physics \\
University of Cambridge, Silver St., Cambridge CB3 9EW}
\date{\today}
\maketitle
\begin{abstract}
Charged dilaton black hole solutions have recently been found for an
action with two $U(1)$ gauge fields and a dilaton field.
I investigate  new exact solutions of this theory analogous to the
C-metric and  Ernst solutions of classical general relativity. The
parameters in the  latter solution may be restricted so that it has
a smooth Euclidean  section with topology $S^2 \times S^2 - \{pt\}$,
which gives an  instanton describing pair production of the charged
dilaton black  holes. These instantons generalize those found
recently by Dowker {\it et al}.
\end{abstract}
\narrowtext

\section{Introduction}
\label{intro}

The study of black hole evaporation explores some of the most
important issues in quantum gravity. One of the most interesting
issues raised by black hole evaporation is the information loss
problem: in the semiclassical theory, the information describing the
configuration of matter used to form the black hole doesn't re-%
emerge in the radiation emitted during collapse, as this
radiation is precisely thermal \cite{Hawk}. It is possible that
higher-order  quantum effects modify the radiation so that it is not
precisely  thermal, but consideration of toy theories in two
dimensions suggest  that such effects become important too late in
the evaporation for  this to resolve the information problem
\cite{CGHS,horns,endpoint}.

Another possible repository for the information lost to the
black hole is some kind of long-lived remnant left behind by the
black hole \cite{horns,rems,comments}. One of the major problems
with this  proposal is that there would have to be an infinite
number of distinct  remnants to account for all the  information
that could possibly have  been dumped into the black hole, as we can
form black holes from  arbitrarily large amounts of matter. One
would na\"{\i}vely assume  that each of these species would have a
finite probability to be pair  produced in a suitable background
field, and there would therefore  be problems with divergences in
the total pair production of  remnants \cite{constraints,infty}.
Various suppression mechanisms  have been suggested which could
produce a finite answer despite  this na\"{\i}ve argument, but they
are the subject of extensive  contention
\cite{rems,comments,constraints,infty}, so an explicit  calculation
of this rate is essential to the further consideration of the
specifically, if black holes can form but never disappear, they
will  violate $CPT$ \cite{thunder}, but I will not concern myself
with this  here.

For a neutral black hole (which the semiclassical theory
predicts will  evaporate completely), the endpoint of evaporation
lies deep in the  quantum regime, as the mass of the black  hole
becomes of order  $M_{pl}$, and it is therefore inaccessible to
semiclassical analysis.  This problem has led to extensive interest
in  consideration of the  quantum behavior of more complicated
black holes \cite{Gary,garfinkle,gidds,Kalloshcens,Kalloshaxion}.
The behavior of  near-extreme charged black holes displays the same
features and  puzzles, and can be studied semiclassically with a
fair degree of  confidence. More complicated models have other
advantages: in  string  theory, the singularity in the sigma-model
metric disappears  from the  spacetime down an infinitely long tube
as extremality is  approached, and  excitations living far down the
throat are   candidate remnants \cite{horns}. It is useful to
consider a model with  no charged particles,  as we then have the
considerable simplification  that the charge of the black hole is
constant, while the fundamental  black hole physics remains
unchanged.

The most common such model is the action
\begin{equation} \label{Faction}
S = \int d^4 x \sqrt{-g} ( R - 2 \partial^{\mu}\phi \partial_{\mu}
\phi -e^{-2a\phi} F_{\mu\nu}F^{\mu\nu} ).
\end{equation}
This action has been extensively considered, and in \cite{Fay},
instantons describing the pair production of black holes in this
theory were developed. In the sum-over-histories approach to
quantum mechanics, the action for these instantons gives a good
approximation to the rate for pair production. Unfortunately, the
instantons are regular only for $a<1$ unless the black
holes produced  are extreme. As the action (\ref{Faction}) for $a=1$
is a part of the  action for the low-energy limit of string theory,
we would like to  extend  consideration to this case.

It has recently been suggested \cite{Kalloshcens} that a
particularly interesting generalization of (\ref{Faction}) is an
action with an additional gauge field,
\begin{eqnarray} \label{Kaction}
I_{SU(4)} &=& \int d^4
x \sqrt{-g} ( R - 2 \partial^{\mu}\phi \partial_{\mu}
\phi  \\
&&-e^{-2\phi} (F_{\mu\nu}F^{\mu\nu} + G_{\mu\nu}G^{\mu\nu})), \nonumber
\end{eqnarray}
where
\begin{equation}
F_{\mu\nu} = \partial_{[\mu}A_{\nu]},\, G_{\mu\nu}
= \partial_{[\mu}B_{\nu]}. \end{equation}
This is a special case of a more general theory with a rigid $SU(4)
\otimes SU(1,1)$ symmetry, vector fields transforming under a
$SU(2) \otimes SU(2)$ group and a complex scalar, which arises from
dimensionally reduced superstring theory or $N=4$ supergravity
\cite{cremmer}. The action (\ref{Kaction}) is invariant under a
duality  transformation,
\begin{mathletters} \label{dualtransf}
\begin{equation}
F_{\mu\nu} \rightarrow \tilde{F}_{\mu\nu} =  \frac{1}{2}  e^{-2\phi}
\epsilon_{\mu\nu\rho\sigma} F^{\rho\sigma},
\end{equation}
\begin{equation}
 G_{\mu\nu} \rightarrow
\tilde{G}_{\mu\nu} =  \frac{1}{2}
e^{-2\phi} \epsilon_{\mu\nu\rho\sigma} G^{\rho\sigma},\, \phi
\rightarrow -\phi,
\end{equation}
\end{mathletters}
which is analogous to the ordinary electric-magnetic duality
transformation of Einstein-Maxwell theory. If we
set one of the  gauge fields to zero, it becomes the
action (\ref{Faction}) with $a=1$. Black hole solutions of
(\ref{Kaction}) were found by Gibbons \cite{Gary}. I seek to develop
C-metric and Ernst solutions of (\ref{Kaction}) which  will
generalize the solutions of (\ref{Faction}) in \cite{Fay} so as to
give a regular instanton describing the pair production of two
non-extreme black holes connected by a throat for $a=1$.

The organization of the rest of the paper is as follows. In section
\ref{C-M} I  discuss the C-metric and black hole solutions of
(\ref{Kaction}), and  relate them to the solutions of
(\ref{Faction}) discussed in \cite{Gary,Fay}.  There are nodal
singularities in the C-metric which cannot in general  be removed by
any choice of the period of the azimuthal coordinate.  In section \ref{Esol}
I describe the appropriate Harrison  transformations to eliminate
the nodal singularities. These  transformations give solutions
analogous to the Ernst solutions, which  reduce to a solution with
two constant background fields when the  size of the black holes is
much less than the scale of the fields, and at  spatial infinity. In
section \ref{inst}, I discuss the Euclidean continuation of  these solutions,
and find that it is necessary to impose another  condition on the
parameters to obtain a regular solution. The regular  instantons
have the topology of $S^2 \times S^2 - \{pt\}$, and thus  describe
pair production of the black holes in the presence of the
background fields \cite{garfinkle}. In the limit that one of the
charges vanishes, the condition on the parameters becomes the
condition for the black holes to be extreme \cite{Fay}. Section \ref{concl}
summarizes my results.

\section{C-Metric Solutions}
\label{C-M}

The charged black hole solutions of (\ref{Kaction}) are
\cite{Gary}:
\begin{equation} \begin{array}{c}
ds^2 = -\lambda dt^2 + \lambda^{-1} dr^2 + R^2 d\Omega,  \\
e^{2\phi} = e^{2\phi_0} \frac{r+\Sigma}{r-\Sigma}, \label{Kbhole} \\
F = \frac{Qe^{\phi_0}}{(r-\Sigma)^2} dt \wedge dr,\; G
=Pe^{\phi_0}\sin \theta d\theta \wedge d\varphi, \end{array}
\end{equation}
where
\begin{equation} \label{Klambda}
\lambda = \frac{(r-r_+)(r-r_-)}{R^2}, \; R^2 = r^2 -\Sigma^2,
\end{equation}
and \cite{stringbh}
\begin{equation} \label{Kparam}
r_{\pm} = M \pm \sqrt{M^2+\Sigma^2-P^2-Q^2},\; \Sigma = \frac{P^2-
Q^2}{2M}.
\end{equation}
There is a curvature singularity at $r=|\Sigma|$. The
physical degrees  of freedom are $P, Q, M$ and $\phi_0$; $M$ is the
mass of the black  hole, $e^{\phi_0} Q$ is its electric charge, and
$e^{\phi_0} P$ is its  magnetic charge. I could keep the asymptotic
value of the dilaton $\phi_0$ as a free parameter, but I will
instead fix it by requiring  that the dilaton match to an
appropriate background value at  infinity. The solution has a
manifest dual symmetry, under
\begin{equation} \label{bhsymm}
Q \leftrightarrow P,\; \Sigma \leftrightarrow -\Sigma,\; \tilde{F}
\leftrightarrow G,\; \phi \leftrightarrow -\phi,
\end{equation}
corresponding to the general symmetry  (\ref{dualtransf}). The
parameters are constrained to $M \geq M_{extr} = (|P|+|Q|)/\sqrt{2}$
by positivity bounds arising from supersymmetry \cite{Kalloshcens}.
The black holes have one unbroken $N=1$ supersymmetry in the
extremal limit if both gauge charges are non-zero, and two if one of
the gauge charges is zero. Unlike the dilaton black hole solutions
of  (\ref{Faction}), where the temperature at infinity is
ill-defined, the temperature of these black  holes goes smoothly to
zero as extremality is approached \cite{Gary,Kalloshcens}.

If we make the coordinate transformation $r' = r+\Sigma$, this
metric becomes
\begin{eqnarray}
&ds^2 = -\lambda dt^2 + \lambda^{-1} dr^2 + R^2 d\Omega,& \nonumber \\
&e^{-2\phi} = e^{-2\phi_0} (1-2\Sigma/r'), & \nonumber \\
&F = \frac{Qe^{\phi_0}}{(r'-2\Sigma)^2} dt \wedge dr,\; G
=Pe^{\phi_0}\sin \theta d\theta \wedge d\varphi,& \label{bhcoord} \\
& \lambda =(r'-r_+ -\Sigma)(r'-r_- - \Sigma)/R^2,& \nonumber \\
& R^2 = r'(r'-
2\Sigma),& \nonumber
\end{eqnarray}
and when $Q=0$ (which implies $r_- = \Sigma = P^2/2M$), this
reduces to the black hole solution of (\ref{Faction}) with magnetic
charge found in \cite{Gary} (and independently in
\cite{stringbh}) if we identify $r'_{\pm} = r_{\pm} +  \Sigma,\,
q=P$. Similarly, if we make a coordinate transformation
$r'=r-\Sigma$ in (\ref{Kbhole}), we will see that it will reduce to
the  electrically charged dual solution \cite{Gary}  when $P=0$ if
we identify $r'_{\pm} =  r_{\pm} - \Sigma,\, q=Q$.

The generalization of the C-metric solution \cite{Cmetric} to this
theory is given by
\begin{eqnarray}
ds^2 = \frac{1}{A^2(x-y)^2}[F(x)(G(y)dt^2-G^{-1}(y)dy^2)
\label{Cmetric} \\
 +F(y)(G^{-1}(x) dx^2 + G(x)d\varphi^2)], \nonumber
\end{eqnarray}
\begin{equation} \label{Cmdil}
 e^{-2\phi} = e^{-2\phi_0} \left(\frac{1+\Sigma A y}{1- \Sigma
Ay}\right) \left(\frac{1-\Sigma Ax}{1+\Sigma A x} \right),
\end{equation}
\begin{equation} \label{Cmgauge}
F_{yt} = \frac{\alpha e^{\phi_0}}{(1+\Sigma Ay)^2},\;
G_{x\varphi}=  \frac{\beta e^{\phi_0}}{(1-\Sigma Ax )^2},
\end{equation}
where
\begin{equation} \label{Cmfnsf}
F(\xi) = 1-\Sigma^2 A^2 \xi^2,
\end{equation}
\begin{equation} \label{Cmfnsg}
 G(\xi) = \frac{(1-\xi^2-r_+ A
\xi^3)(1+r_- A\xi)}{(1- \Sigma^2 A^2 \xi^2)},
\end{equation}
and
\begin{eqnarray} \label{alpha}
\alpha^2 &=& \frac{1}{2}  (r_+ -\Sigma)(r_- -\Sigma) +\frac{1}{2}  A^2
\Sigma^3 (r_-  -\Sigma)\\
&=& Q^2
+ \frac{1}{2}  A^2 \Sigma^3 (r_- -
\Sigma),  \nonumber
\end{eqnarray}
\begin{eqnarray} \label{beta}
\beta^2 &=& \frac{1}{2}  (r_+ + \Sigma)(r_- + \Sigma) - \frac{1}{2}
A^2 \Sigma^3 (r_-  + \Sigma) \\
&=& P^2 -\frac{1}{2}  A^2 \Sigma^3 (r_- +
\Sigma). \nonumber
\end{eqnarray}
This metric has the same general form as the C-metric solutions in
\cite{Fay,Cmetric},  and we choose coordinates so that the cubic
factor in $G$ still has the same form, but the functions $G$ and $F$
are more complicated. The parameters in this solution are still
related by (\ref{Kparam}). The interpretation of  the  parameters is
now essentially qualitative, as the mass and charges are given by $M,
Q$ and $P$ only in the weak-field limit. The additional  parameter
$A$ determines the strength of the acceleration. Note that this
solution also has the manifest dual symmetry  (\ref{bhsymm}).

This C-metric solution tends to the charged black hole
(\ref{Kbhole})  as $A \rightarrow 0$. To see this, make the
coordinate  transformation
\begin{equation} \label{Atransf}
r = -\frac{1}{Ay},\, T = A^{-1}t,
\end{equation}
which puts the metric in the form
\begin{eqnarray}
ds^2 = \frac{1}{(1+Arx)^2}[F(x)(-H(r)dT^2+H^{-1}(r)dr^2)
\label{Ametric} \\
 +R^2(r)(G^{-1}(x)dx^2+G(x)d\varphi^2)], \nonumber
\end{eqnarray}
where
\begin{equation}
H(r) = \frac{(r-r_+-A^2r^3)(r-r_-)}{R^2(r)},
\end{equation}
$F(x)$ is given by (\ref{Cmfnsf}), and $R(r)$ is given by
(\ref{Klambda}). If we now set $A=0$ and make the further
coordinate transformation $x=\cos \theta$, this metric reduces to
the  charged black hole metric (\ref{Kbhole}). The other fields also
reduce  to the forms in (\ref{Kbhole}). When $\Sigma=0$ ({\it i.e.},
$P=Q$), the metric reduces to the charged C-metric of
Einstein-Maxwell theory  \cite{Cmetric}, but with a different field
content.

If either of the gauge charges vanish, this C-metric will reduce
to the  dilaton C-metric given in \cite{Fay} with $a=1$, although
the  parameters are not simply related to the parameters $m,A,q$ of
that  solution. To demonstrate this in the case where $Q$ is set to
zero, we  first perform a coordinate transformation
\begin{equation} \label{Qtransf1}
y = \frac{y'}{1+\Sigma A y'},\, x = \frac{x'}{1 + \Sigma A x'};
\end{equation}
the metric (\ref{Cmetric}) becomes
\begin{eqnarray} ds^2 = \frac{1}{A^2(x'-y')^2}[F(x')(G(y')dt^2-G^{-
1}(y')dy'^2) \label{Qmetric} \\
+F(y')(G^{-1}(x') dx'^2 + G(x')d\varphi^2)], \nonumber
\end{eqnarray}
where
\begin{equation}
F(\xi') = 1+2\Sigma A \xi',
\end{equation}
\begin{equation}
 G(\xi') = \frac{[(1+\Sigma A \xi')^3
- \xi'^2  -r_+'A \xi'^3](1+r_-' A \xi')}{F(\xi')},
\end{equation}
and $r'_{\pm} = r_{\pm} + \Sigma$. If we now make a further
coordinate transformation \cite{Cmetric}
\begin{equation} \label{Qtransf2}
t = c_0 \hat{t},\, \varphi = c_0 \hat{\varphi},\, x' = c_1
c_0 \hat{x} + c_2,\,  y'=c_1c_0 \hat{y} + c_2,
\end{equation}
and restrict $c_0, c_1, c_2$ suitably, we can rewrite this as
\begin{eqnarray}
ds^2 = \frac{1}{\hat{A}^2(\hat{x}-\hat{y})^2}  [\hat{F}(\hat{x})
(\hat{G}(\hat{y})d\hat{t}^2- \hat{G}^{-1}(\hat{y}) d\hat{y}^2)
\label{Qmetric2} \\
 +\hat{F}(\hat{y})(\hat{G}^{-1}(\hat{x}) d\hat{x}^2 + \hat{G}(\hat{x})
d\hat{\varphi}^2)], \nonumber
\end{eqnarray}
where now
\begin{equation}
\hat{F}(\hat{\xi}) = 1+2\hat{\Sigma}\hat{A}\hat{\xi},
\end{equation}
 and
\begin{equation}
\hat{G}(\hat{\xi}) =
\frac{(1-\hat{\xi}^2- \hat{r}_+\hat{A} \hat{\xi}^3)(1+\hat{r}_-
\hat{A}
\hat{\xi}) }{\hat{F}(\hat{\xi})}.
\end{equation}
The dilaton and gauge fields in this coordinate system are
\begin{equation}
e^{-2\phi} = e^{-2\phi_0} \frac{\hat{F}(\hat{y})}{\hat{F}(\hat{x})},
\end{equation}
\begin{equation}
G_{\hat{x}\hat{\varphi}} = \beta e^{\phi_0},\text{ and }
 F_{\hat{y}\hat{t}} =
\frac{\alpha e^{\phi_0}}{\hat{F}^2(\hat{y})},
\end{equation}
where we find
\begin{equation}
\alpha^2 = \frac{1}{2} (\hat{r}_+
-2\hat{\Sigma})(\hat{r}_- -2\hat{\Sigma})  + \frac{1}{2}
\hat{\Sigma}^3 \hat{A}^2 (\hat{r}_- - 2\hat{\Sigma})
\end{equation}
and
\begin{equation}
\beta^2 =  \frac{1}{2}  \hat{r}_+ \hat{r}_-.
\end{equation}
The parameters are related by
\begin{equation}
\hat{A}^2 = \frac{A^2 c_1}{1+2\Sigma A c_2},
\end{equation}
\begin{equation} \label{rel}
\hat{r}_-\hat{A} =
\frac{r'_- A}{1+2\Sigma A c_2},\, \hat{\Sigma}\hat{A} = \frac{\Sigma
A}{1+2\Sigma A c_2},
\end{equation}
and
\begin{equation}
\hat{r}_+ \hat{A} = r'_+ A c_0^3 c_1^2 - \Sigma^3 A^3 c_0^3 c_1^2.
\end{equation}
When $Q=0$, $r_-' = 2\Sigma$, and (\ref{rel}) therefore implies
$\hat{r}_- =2\hat{\Sigma}$. Thus,  (\ref{Qmetric2}) then reduces to
the C-metric solution given in \cite{Fay} for $a=1$. When $P=0$, a
similar transformation may be used to show that it  reduces to the
electric dual to the $a=1$ solution in \cite{Fay}.

For $r_+ A < 2/(3\sqrt{3})$, the function $G(\xi)$ has four
real roots,  which we denote in ascending order by $\xi_1,\,
\xi_2,\, \xi_3,\,  \xi_4$. We may restrict the parameters so that
$\xi_1 = -1/r_- A$ and $\xi_1 < \xi_2 \leq \xi_3 < \xi_4$.
The surface $y = \xi_0  \equiv -1/|\Sigma| A$ is singular;
this surface is analogous to the singular surface at $r=|\Sigma|$ in
the black hole solutions  (\ref{Kbhole}). As $r_- \geq |\Sigma|,\,
\xi_1 \geq \xi_0$. The surfaces  $y=\xi_1,\, y=\xi_2$ are the black
hole horizons, and $y=\xi_3$ is the  acceleration horizon. If I
allowed $\xi_1 = \xi_2$, the black hole  would be extremal, although
the horizons would still be regular.  However, for this paper, I will
restrict my attention to $\xi_1 <  \xi_2$. The coordinates
$(x,\varphi)$ are angular coordinates, and  $x$ is restricted to the
range $\xi_3 \leq x \leq \xi_4$ in which  $G(x)$ is positive, so
that the metric has the appropriate signature.  At $x=\{\xi_3 ,\,
\xi_4\}$, the norm of $\partial/\partial \varphi$ vanishes, so  these points
are interpreted as the poles of two-spheres around the  black hole.
There is a divergence in the metric at $x=y$, which is  interpreted
as the point at infinity, so $y$ is restricted to the range
$\xi_0 < y < x$. Spatial infinity is reached when $y=x=\xi_3$,
and  null or timelike infinity when $y=x\neq \xi_3$ \cite{ashtekar}.

It is not generally possible to choose the period $\Delta \varphi$
of  the azimuthal coordinate so that the nodal singularities at the
two  poles $x = \xi_3$ and $x=\xi_4$ are eliminated simultaneously.
The  deficit angles at the two poles are given by \cite{Fay}
\begin{equation}
\label{deficit} \delta_3 = 2\pi -\frac{1}{2} \Delta \varphi
|G'(\xi_3)|,\, \delta_4 = 2\pi -  \frac{1}{2}  \Delta \varphi
|G'(\xi_4)|. \end{equation}
By choosing $\Delta \varphi = 4\pi/|G'(\xi_3)|$ we may eliminate the
nodal singularity at $x=\xi_3$, but there will then in general be a
negative deficit angle along the $\xi_4$ direction, which may be
interpreted as a ``line singularity'' pushing the two black
holes apart.  In the next section, we will see how we can eliminate
the nodal  singularities by introducing external fields via a
Harrison  transformation \cite{Htransf}.

\section{Ernst Solutions}
\label{Esol}

In \cite{Fay}, Dowker {\it et al} showed that given an axisymmetric
solution with $A_i = g_{i\varphi} =0$, where $x^i$ are the
other three  coordinates, a new axisymmetric solution may be
obtained by the  transformation
\begin{equation}
g'_{ij} = \Lambda g_{ij},\, g'_{\varphi\varphi} = \Lambda^{-1}
g_{\varphi\varphi},
\end{equation}
\begin{equation}
e^{-2\phi'} = e^{-2\phi} \Lambda,\, A'_{\varphi} =
-\frac{1}{B\Lambda}(1+B A_{\varphi}), \label{EHtransf}
\end{equation}
\begin{equation}
\Lambda = (1+B A_{\varphi})^2+ \frac{1}{2}  B^2 g_{\varphi\varphi}
e^{2\phi}.
\end{equation}
I will now construct an appropriate generalization of this
transformation. In the string conformal gauge $ds_T^2 = e^{2\phi}
ds^2$, the action (\ref{Kaction}) is
\begin{equation}
S = \int d^4 x \sqrt{-g_T} e^{-2\phi} \left[ R_T + 4 (\nabla
\phi)^2 -  F^2 -G^2 \right].
\end{equation}
If I write $F_{\mu\nu}$ and $G_{\mu\nu}$ in terms of vector
potentials  as
\begin{equation}
G_{\mu\nu} = \partial_{[\mu} A_{\nu]},\, F_{\mu\nu} = \frac{1}{2}
 e^{2\phi}  \epsilon_{\mu\nu\rho\sigma} \partial^\rho B^\sigma
\end{equation}
and introduce the definitions
\begin{equation}
\,^3 \! g_{ij} = g_{Tij},\, V = g_{T\varphi\varphi},\, \tilde{\phi}
= \phi -  \frac{1}{4} \log V,
\end{equation}
then the action can be rewritten in the form
\begin{eqnarray}
S &=& \alpha \int d^3 x \sqrt{-^3\! g}e^{-2\tilde{\phi}} \left[ \,^3 \!
R + 4 \partial_i  \tilde{\phi} \partial^i \tilde{\phi} \right. \\
&& - \frac{1}{4}
V^{-2} \partial_i V \partial^i  V -2 V^{-1}
\partial_i A_{\varphi} \partial^i A_{\varphi} \nonumber \\
&&\left. -2 e^{4\tilde{\phi}} \partial_i B_{\varphi}  \partial^i
B_{\varphi} \right].  \nonumber
\end{eqnarray}
It is now a relatively easy exercise to show that this action is
invariant under the transformations
\begin{equation}
 V' = \frac{1}{\Lambda^2} V,\,  A_{\varphi}' = -\frac{1}{B\Lambda}
(1+B A_\varphi),
\end{equation}
\begin{equation}
\,^3 \! g_{ij}' = \Psi^2\, \,^3 \! g_{ij},\, e^{-2\tilde{\phi'}}
= \frac{1}{\Psi} e^{-2\tilde{\phi}},
\end{equation}
\begin{equation}
  B'_\varphi =
-\frac{1}{E\Psi}(1+ E B_\varphi),
\end{equation}
\begin{equation}
 \Lambda = (1+B A_\varphi)^2 + \frac{1}{2}  B^2 V,
\end{equation}
\begin{equation}
 \Psi = (1+E B_\varphi)^2 + \frac{1}{2}  E^2 e^{-4\tilde{\phi}}.
\end{equation}
I may now construct new solutions by applying these
transformations to axisymmetric solutions satisfying $A_i =
g_{i\varphi} = 0$. The analogue of the Melvin
solution \cite{melvin},  obtained by applying these transformations
to the vacuum,  is\footnote{After we have made a further gauge
transformation to  make the vector potentials regular on the axis
$\rho=0$.}
\begin{equation}
\label{dualMelvin} ds^2 = \Lambda \Psi[-dt^2+d\rho^2+dz^2] +
\frac{\rho^2 d\varphi^2}{\Lambda \Psi},
\end{equation}
\begin{equation} \label{dMgauge}
e^{-2\phi} = \frac{\Lambda}{\Psi},\, A_\varphi =
-\frac{B\rho^2}{2\Lambda},\,
B_\varphi = -\frac{E\rho^2}{2\Psi},
\end{equation}
\begin{equation}
G_{\mu\nu} = \partial_{[\mu} A_{\nu]},\, F^{\mu\nu} = \frac{1}{2}
 e^{2\phi}  \epsilon^{\mu\nu\rho\sigma} \partial_{\rho} B_{\sigma},
\end{equation}
\begin{equation}
\Lambda = 1+ \frac{1}{2}  B^2 \rho^2,\, \Psi = 1+\frac{1}{2} E^2
\rho^2. \end{equation}
This solution has a manifest dual symmetry
\begin{equation}
B \leftrightarrow E,\, \tilde{F} \leftrightarrow G,\,
\phi \leftrightarrow  -\phi.
\end{equation}

If we apply the transformations to (\ref{Cmetric}), we will obtain
an  analogue of the Ernst solution \cite{Ernst} which preserves the
manifest dual symmetry of (\ref{Cmetric}). The resulting solution is
\begin{eqnarray}
ds^2 &=& \frac{\Lambda\Psi}{A^2(x-y)^2}[F(x)(G(y)dt^2-G^{-1}(y)
dy^2) \label{Ernst} \\
&& +F(y)G^{-1}(x) dx^2]+ \frac{F(y)G(x)}{\Lambda\Psi A^2(x-y)^2} d\varphi^2,
\nonumber
\end{eqnarray}
\begin{equation} \label{Ernstf}
e^{-2\phi} = e^{-2\phi_0} \frac{\Lambda}{\Psi} \left(\frac{1+\Sigma
A y}{1- \Sigma Ay}\right) \left(\frac{1-\Sigma Ax}{1+\Sigma A x}
\right),
\end{equation}
\begin{equation}
A_{\varphi} = -\frac{e^{\phi_0}}{B\Lambda}\left(1+\frac{B\beta
x}{1-
\Sigma Ax}\right)+k,
\end{equation}
\begin{equation}
B_{\varphi} = -\frac{e^{-\phi_0}}{E \Psi}
\left(1+ \frac{E \alpha x}{1 + \Sigma Ax}\right)+k',
\end{equation}
\begin{equation}
G_{\mu\nu} = \partial_{[\mu} A_{\nu]},\, F^{\mu\nu} = \frac{1}{2}
e^{2\phi}  \epsilon^{\mu\nu\rho\sigma} \partial_{\rho} B_{\sigma},
\end{equation}
\begin{eqnarray}
\Lambda &=& \left(1+\frac{B\beta x}{1-\Sigma Ax}\right)^2  \\
&& +\frac{B^2(1-
x^2-r_+ A x^3)(1+r_- A x)(1-\Sigma Ay)^2}{2A^2(x-y)^2(1-\Sigma A
x)^2}, \nonumber
\end{eqnarray}
\begin{eqnarray}
\Psi &=& \left(1+\frac{E\alpha x}{1+\Sigma Ax}\right)^2 \\
&&+ \frac{E^2(1-
x^2-r_+ A x^3)(1+r_- A x)(1+\Sigma Ay)^2}{2A^2 (x-y)^2(1+\Sigma A
x)^2}, \nonumber
\end{eqnarray}
where $F(\xi)$ and $G(\xi)$ are given by (\ref{Cmfnsf},\ref{Cmfnsg}),
and $\alpha$
and $\beta$ are given by (\ref{alpha},\ref{beta}). The constants
$\phi_0, k$, and $k'$ will be chosen so that the solution
at infinity  agrees with (\ref{dualMelvin}). This solution has the
manifest dual  symmetry
\begin{equation} \label {Ernstsym}
Q \leftrightarrow P,\, \Sigma \leftrightarrow -\Sigma,\, B
\leftrightarrow E,\, \tilde{F} \leftrightarrow G,\, \phi \leftrightarrow
-\phi
\end{equation}
(which implies $k \leftrightarrow k'$ and $\phi_0 \leftrightarrow
-\phi_0$ to preserve the agreement with (\ref{dualMelvin}) at
infinity).

As in the original Ernst solution, the background fields provide the
force necessary to accelerate the black holes. To eliminate
the nodal  singularities in this metric at $x=\xi_3$ and $x=\xi_4$
simultaneously, we must constrain $B$ and $E$ so that
\begin{equation} \label{constraint2}
G'(\xi_3) \Lambda(\xi_4) \Psi(\xi_4) = -G'(\xi_4)
\Lambda(\xi_3) \Psi(\xi_3)
\end{equation}
 and choose $\Delta \varphi=4\pi|\Lambda
\Psi/G'(x)|_{x=\xi_4}$\footnote{Note that $\Lambda(\xi_i) \equiv
\Lambda(x=\xi_i)$ and $\Psi(\xi_i) \equiv
\Psi(x=\xi_i)$ are constants.}. In  the limit $r_+ A \ll
1$ \cite{Fay},  this constraint reduces to Newton's law,
\begin{equation}
MA \approx BP + EQ.
\end{equation}
This leads me to suppose that $r_+ A \ll 1$ is in some sense a point
particle limit, which appears reasonable, as this is simply
a statement  that the black hole is small on the scale set by the
background fields.  In this limit, so long as $|r_+ A y| \ll 1$ as
well, one finds that $G(\xi)  \approx 1-\xi^2, F(\xi) \approx 1$
\cite{Fay}, and thus the metric  (\ref{Ernst}) becomes
\begin{eqnarray}
ds^2 &\approx & \frac{\Lambda\Psi}{A^2(x-y)^2}[(1-y^2)dt^2-(1-
y^2)^{-1} dy^2\\
&& +(1-x^2)^{-1} dx^2]
+ \frac{1-x^2}{\Lambda\Psi A^2(x-y)^2} d\varphi^2, \nonumber
\end{eqnarray}
\begin{equation}
\Lambda \approx 1+ \frac{1}{2}  B^2 \frac{1-x^2}{A^2(x-y)^2},
\end{equation}
\begin{equation}
\Psi \approx  1+ \frac{1}{2}  E^2 \frac{1-x^2}{A^2(x-y)^2}.
\end{equation}
This is just the Melvin solution (\ref{dualMelvin}) in non-standard
coordinates: the transformation
\begin{equation}
\rho^2 = \frac{1-x^2}{A^2(x-y)^2},\, \zeta^2 =
\frac{y^2-1}{A^2(x-y)^2}, \end{equation}
\begin{equation}
\hat{t} = \zeta \sinh t,\, z = \zeta \cosh t,
\end{equation}
puts it in the form (\ref{dualMelvin}). The dilaton and gauge fields
(\ref{Ernstf}) in this approximation are
\begin{equation}
e^{-2\phi} \approx e^{-2\phi_0} \frac{\Lambda}{\Psi},\, A_{\varphi}
\approx \frac{e^{\phi_0} B \rho^2}{2 \Lambda},\, B_{\varphi}
\approx \frac{e^{-\phi_0}E \rho^2}{2 \Psi},
\end{equation}
where $k$ and $k'$ have been chosen so as to give regularity on the
axis $\rho=0$, in agreement with (\ref{dMgauge}). This agrees with
(\ref{dMgauge}) up to the arbitrary constant shift of the dilaton.

The Ernst solution (\ref{Ernst}) also approaches (\ref{dualMelvin})
at  large spacelike distances. Spatial infinity corresponds to $x, y
\rightarrow \xi_3$, and in this limit it is convenient to use the
change of coordinates given in \cite{Fay},
\begin{equation} \label{inftransf}
x - \xi_3 = \frac{4 F(\xi_3)
L^2}{G'(\xi_3)A^2}\frac{\rho^2}{(\rho^2+\zeta^2)^2},
\end{equation}
\begin{equation}
\xi_3 -y =
\frac{4 F(\xi_3)L^2}{G'(\xi_3) A^2}
\frac{\zeta^2}{(\rho^2+\zeta^2)^2}, \end{equation}
\begin{equation}
t= \frac{2 \eta}{G'(\xi_3)},\, \varphi = \frac{2L^2
\tilde{\varphi}}{G'(\xi_3)},
\end{equation}
where I have introduced $L^2 = \Lambda(x=\xi_3) \Psi(x=\xi_3)$.  Note
that the choice of period of $\varphi$ implies $\tilde{\varphi}$
has period $2\pi$. For large $\rho^2+\zeta^2$, the Ernst solution
in  these coordinates reduces to \begin{equation}
ds^2 \rightarrow \tilde{\Lambda}\tilde{\Psi} (-\zeta^2 d\eta^2 +
d\zeta^2 + d\rho^2) + \frac{\rho^2
d\tilde{\varphi}^2}{\tilde{\Lambda}\tilde{\Psi}},
\end{equation}
where
\begin{equation}
\tilde{\Lambda} = (1+\frac{1}{2}  \hat{B}^2 \rho^2) \text{ with
} \hat{B}^2  =\frac{B^2  G'^2(\xi_3)}{4L^2\Lambda(\xi_3)},
\end{equation}
and
\begin{equation}
\tilde{\Psi} = (1+\frac{1}{2}  \hat{E}^2 \rho^2) \text{ with
} \hat{E}^2 =  \frac{E^2 G'^2(\xi_3)}{4L^2\Psi(\xi_3)}.
\end{equation}
If we now set $\hat{t} = \zeta \sinh \eta, z = \zeta \cosh \eta$, we
once again regain (\ref{dualMelvin}). For large $\rho^2+\zeta^2$,
the  dilaton and gauge fields tend to
\begin{equation}
e^{-2\phi} \rightarrow L^2 e^{-2\phi_0}
\frac{\tilde{\Lambda}}{\tilde{\Psi}},
\end{equation}
\begin{equation}
A_{\tilde{\varphi}} \rightarrow L^{-1} e^{\phi_0}
\frac{\hat{B}\rho^2}{ 2\tilde{\Lambda}},\, B_{\tilde{\varphi}}
\rightarrow L e^{-\phi_0} \frac{\hat{E} \rho^2}{2\tilde{\Psi}},
\end{equation}
so if we set $e^{\phi_0}=L$, we recover (\ref{dMgauge}) in this
limit. I  will take this to define $\phi_0$ in general. In summary,
we recover  the Melvin solution at large spacelike distances, with
the physical  background fields $\hat{E}$ and $\hat{B}$.  In the
limit  $r_+ A \ll 1,  \hat{B} \approx B, \hat{E} \approx E$, as
expected.

\section{Instantons}
\label{inst}

The solution (\ref{Ernst}) describes two black holes accelerating
away  from each other, propelled by the constant background fields.
I now  consider the Euclidean section obtained by taking $\tau = it$
in  (\ref{Ernst}). The Euclidean section gives an exact instanton
describing pair production of the accelerating black holes
\cite{karpacz,garfinkle}. I will only consider the non-extreme or
wormhole  instantons, {\it i.e.}, $\xi_1<\xi_2$. I then find that it
is necessary to  impose another condition on the parameters to
eliminate the possible  conical singularities at the black hole
horizon $y=\xi_2$ and the  acceleration horizon $y=\xi_3$
simultaneously. Namely, we must take the  period of $\tau$ to be $\Delta
\tau = 4\pi/G'(\xi_2)$ and set
\begin{equation}
|G'(\xi_2)| = |G'(\xi_3)|,
\end{equation}
where $G(\xi)$ is given by (\ref{Cmfnsg}).
This condition may be satisfied in either of two ways. Firstly,
we may  set $\xi_3=\xi_2$, which gives a regular instanton with
topology $S^2  \times R^2$, whose physical interpretation is unclear
\cite{Fay}.  Alternatively, we may set
\begin{equation}
\left(\frac{\xi_2^2-\xi_0^2}{\xi_3^2-\xi_0^2}\right) \left(
\frac{\xi_3-\xi_1}{\xi_2-\xi_1} \right)= \frac{\xi_4-\xi_2}{\xi_4 -
\xi_3}. \label{instcond}
\end{equation}
This condition provides a further restriction on the four parameters $Q$,
$P$, $M$ and $A$, which it is useful to think of as determining $r_+A$ in terms
of $r_-A$ and $\Sigma A$. It is difficult to give the explicit solution, as the
roots $\xi_2,\ \xi_3$, and $\xi_4$ are complicated functions of $r_+A$.
We can however make some interesting general remarks about the solution.

The left-hand side of (\ref{instcond}) is greater than one,
so the right-hand side must be greater than one as well. Since the
first factor on the right-hand side is less than one, this requires that
the second factor be significantly greater than one.
Therefore, the condition will only be satisfied if $\xi_2-\xi_1$
is sufficiently small ({\it i.e.}, for black
holes sufficiently  close to extremality). In the `point-particle'
limit $r_+A\ll 1$, (\ref{instcond}) reduces to $r_+ \approx r_-$. Similarly, in
the limit $Q \rightarrow 0$ (or $P \rightarrow 0$, but not both),
it reduces to $\xi_2 \approx \xi_1$.
Thus, we can only construct a regular instanton for the production of
extreme black holes in these two limits. This asymptotic behavior
is comparable to that of the condition on the wormhole instantons
in \cite{Fay}.

The topology of the instanton is $S^2 \times S^2 - \{pt\}$,
where the  removed point is $x=y=\xi_3$. These instantons are
therefore a  suitable generalization of the wormhole instantons of
\cite{Fay}. To  see that they can be interpreted as a bounce, note
that the $\tau=0,  \tau=\pi/2$ section has topology $S^2 \times S^1
- \{pt\}$, which is  the topology of a wormhole attached to a
spatial slice of the Melvin  universe (\ref{dualMelvin}). It
describes the  production of a pair of oppositely  charged black
holes (\ref{Kbhole})  connected by a wormhole throat, which
subsequently accelerate  away from each other.

\section{Conclusions}
\label{concl}

We have seen that it is possible to extend the construction of
analogues of the C-metric and Ernst solutions in \cite{Fay} to the
theory with two $U(1)$ gauge fields  \cite{Kalloshcens}. The
resulting  solutions share the property of dual symmetry with the
black hole  solution of this theory, (\ref{Kbhole}), and the C-metric
solution  reduces to (\ref{Kbhole}) when the acceleration parameter
$A  \rightarrow 0$. The C-metric solution also reduces to the
dilaton C-metric solutions of \cite{Fay} when one of the gauge
charges vanish,  although the parameters in (\ref{Cmetric}) are not
simply related to  the parameters in the dilaton C-metric solution
of \cite{Fay}.

The instantons discussed here extend the conclusions of
\cite{Fay} to  the case $a=1$. That is, they describe pair creation
of black holes in a  pair of background fields. The fact that this
was possible in the action  with two $U(1)$ gauge fields
(\ref{Kaction}) and not in the previously  considered action
(\ref{Faction}) is related to the thermodynamic  properties of the
black holes (\ref{Kbhole}). In \cite{Kalloshcens}, it  was pointed
out that the puzzling thermodynamic behavior of  dilaton black
holes in the extremal limit may be resolved by  considering a more
general class of black holes, with a dilaton and  two $U(1)$ gauge
fields. The temperature of these black holes goes  smoothly to zero
in the extremal limit, so long as both charges are  non-vanishing.

Well-behaved instantons do exist for $a=1$. This leads me to believe
that it should be possible to study the question of information loss
and related issues in low-energy string theory semiclassically, at
least where the temperature is well-defined. In particular,
remnants  provide a potentially viable solution to the information
loss paradox  even for $a=1$. The instantons presented here also
seem to avoid the  problem of infinite pair production, although it
will be necessary to  calculate their action explicitly, and give a
more careful consideration  to quantum perturbations, before this
problem can be said to be  resolved. These calculations and the
relation of these instantons to  the extremal instantons of
\cite{Fay} will be the subject of a future  paper.

\acknowledgments

I have received many helpful suggestions and comments from Stephen
Hawking in connection with this work.
I gratefully acknowledge the financial support of the Association
of Commonwealth Universities and the Natural Sciences and
Engineering Research Council of Canada.

\end{document}